\documentclass[useAMS,usenatbib,usegraphicx,onecolumn]{mn2e}

\title[Geometry variation of accretion disk during outbursts]{A Study of the Variation of Geometry of Accretion Flows of Compact Objects 
through Timing and Spectral Analysis of Their Outbursts}
\author[P. S. Pal and S. K. Chakrabarti]{P. S. Pal$^{1}$\thanks{E-mail:
parthasarathi.pal@gmail.com; partha.sarathi@bose.res.in} and 
S. K. Chakrabarti$^{1}$\thanks{chakraba@bose.res.in}\\
$^{1}$S. N. Bose National Centre For Basic Sciences, Kolkata, 700098, India\\
}
\begin{document}

\date{}

\pagerange{\pageref{firstpage}--\pageref{lastpage}} \pubyear{2014}

\maketitle

\label{firstpage}

\begin{abstract}

Temporal and spectral variations of black hole candidates during outbursts have been 
reported in several publications. It is well known that during an outburst, source becomes 
softer in first few days, and then returns to hard states after a few weeks or months. In 
the present paper, we show variation of Comptonization Efficiency (CE), defined to be the
ratio of number of power-law photons and number of soft photons injected to into the Compton cloud,
as a function of time in several outbursts. Since power-law photons are generated through 
inverse-Comptonization of the intercepted soft photons, CE is a measure of 
geometry of Compton cloud. Our investigation indicates that all outbursts start
with a large CE and becomes very small after a few days, 
when the Compton cloud becomes small enough to intercept any significant 
number of soft photons. CE returns back to a larger value at the end of outburst. 
We show co-variation of count rates, frequency of quasi-periodic oscillations (QPOs), 
photon index and CE and establish a general trend of disk geometry variation 
during outbursts in all systems under consideration.

\end{abstract}


\section{Introduction}

It is well known that black hole accretion flows in a compact binary system    
typically consist of a Keplerian disk which emits soft or low energy 
photons and a hot Compton cloud which inverse Comptonizes soft
photons into high energy photons \citep{sun85}. When a black hole spectral state
changes from hard to soft and vice versa, 
Compton cloud must change its shape and temperature \citep{C95,t95}: 
In a soft state, when the accretion rate in the Keplerian disk is high, 
Compton cloud is smaller and cooler, while in a hard state, when Keplerian disk rate is very low, 
Compton cloud is larger and hotter. There are various models of the 
Compton cloud in the literature, ranging from a hot Corona, magnetic corona, 
to post-shock region of an accreting low angular momentum flow which surrounds the standard
Kelerian disk \citep{H91, w91, e97, J00, MF01, z03, N11, O05, li05, C95, CM00, rao00, M06}. 
Along with spectral variabilities, a black hole candidate shows temporal variations also,
which is sometimes quasi-periodic \citep{M99, C08, C09}. There are various explanations 
of quasi-periodic oscillations in literature ranging from disc-seismology to oscillation
of post-shock region \citep{k05, ax05, CM00}. 

Outburst sources are ideal candidates to study changes in the size of a Compton cloud. This is
because object is known to be in a hard state at the beginning of an outburst but in a matter
of few days to a few weeks, it changes its states to several other states, thereby
giving us a unique opportunity to study size variation of Compton cloud very easily. 
From the spectral analysis, it appears that following sequence is typically followed by most, if 
not all, outburst sources: hard $\rightarrow$ hard-intermediate $\rightarrow$ soft-intermediate 
$\rightarrow$ soft $\rightarrow$ soft-intermediate $\rightarrow$ hard-intermediate 
$\rightarrow$ hard) \citep{C08, C09, B10, n12}.

Recently, \citet{P11} showed that average size of Compton cloud in 
the variable source GRS 1915+105 changes in a very well defined way
as it transits from one variability class to another. They computed a quantity called 
Comptonizing Efficiency (CE) which is the ratio of number of power-law photons to 
the number of blackbody photons in the spectrum at a given instant. They show that CE is very 
small in softer classes and larger for harder classes. Within some of classes, 
there are evidences of rise and fall of count rates and spectral slopes in a matter of a few seconds. 
Since number of hard photons depends on optical depth of Compton cloud, 
we clearly see a change in optical depth of the Compton cloud in that time scale. 
Furthermore, we see that one variability class goes to another adjacent class whose CE
is nearest.

In the present paper, we study variation of Compton cloud size during outbursts of several 
black hole candidates. We show that outbursts typically start with a large Comptonization efficiency,
i.e., a large sized Compton cloud with a poor soft photon source. 
In rising phase, this cloud becomes progressively smaller and smaller on a daily basis, till 
it became minimum when the object went to a soft state. Of course, soft photon source gets stronger
(soft photon intensity rises) and as a result, the cloud size shrinks (QPO frequency, assumed to be the 
oscillating frequency of the cloud, rises). 
In declining phase of an outburst, the trend is exactly reversed. 
In this paper, we deal with black hole candidates which exhibited outbursts, such as,
H1743-322, GX 339-4, 4U 1543-47, XTE J1118+480, XTE J1859+226, GRO J1655-40 and XTE J1550-564.
In the next Section, we briefly discuss the objects of our study. In \S 3, we present the procedure
of our analysis. Here, we compute Comptonization Efficiency (CE) of all these outbursts both 
in rising and declining phases. We show that CE changes with time very rapidly. 
In general, softer states were found to have a very low CE and the harder states (at the 
beginning of rising phase and at the end of declining phase) 
were found to have a very high CE. In \S 4, we present results and interpret them using 
physical models. Finally, in \S 5, we draw our conclusions.

\section{Objects of our Study}

We consider following outburst sources in the study.

(a) The black hole candidate H1743-322 was first discovered by Ariel in 5 August, 
1977 \cite{k77}. Later, it was precisely located by High Energy 
Astronomical observatory 1 (HEAO) by \citet{D77}. This object was classified 
as a black hole candidate by \citet{w83}. In March 2003, an outburst
along with an increase of source flux by a factor of 3 (60 mCrab in 15-40 keV energy range), 
was reported by INTEGRAL where X-Ray properties of a black hole candidate were 
studied \citep{rv03}. Another outburst, along with increase of source flux
from 16 - 160 mCrab in 2-10 keV intensity,  
was reported by \cite{s04} in July, 2004. Several radio ejections are
reported during the 2003 outburst \citep{r03}. Later, from the 
ejected plasma, radio and X-ray synchrotron emissions are also reported \citep{C05}.
Distance of the object is $8$ kpc \citep{C05} and 
mass is estimated to be $M=10.0 ~ M_\odot$ \citep{C10,Mc09}.
Outburst of 2009 was analyzed by \citet{C10}. 
Outbursts of 2008 and other outbursts are analyzed by \citet{cor11} 
and \citet{J10}. Spectral fitting is done with 
$n_H = 1.6 \times 10^{22} ~ cm^{-2}$ \citep{Cap09}.

%For the analysis purpose we have taken the PCU2 data and considered 1\% systematic error (Mclintock et al. 2009).    

(b) GX 339-4 is a stellar mass Galactic Black hole candidate. This black hole candidate was 
first observed by MIT X-Ray detector of OSO-7 during the survey period
between October, 1971 and January, 1973 between 1-60 keV energy range. This LMXB is 
located at (l,b)=338.93, -4.27 \citep{M73} with RA =  $17^h 02^m 49^s.36$
and dec=$-48^{\circ}47´22¨.8 (J2000)$. Optical observation leads to an estimation of 
mass function to be around $5.8 \pm 0.5 ~ M_{\odot}$ \citep{H03a} and $D=5.8$ kpc \citep{H04}. 
Estimated mass of this object is $7.5 \pm 0.8 ~ M_{\odot}$ \citep{C11}. 
During RXTE {\it Rossi X-Ray Timing Explorer} Era, GX 339-4 had undergone several outburst phases   
(1998, 2002/2003, 2004/2005, 2006/2007, 2010). 
GX 339-4 was observed several times by RXTE during outburst period \citep{D09, M09}.
In 2006, \citet{K06} reported beginning of 2007/2007 outburst of GX 339-4 
with increase of source flux to 230 mCrab in 15-50 keV energy range by Swift-BAT.
In 2010, \citet{Y10} reported an increase in source flux from 17 mCrab to $26 \pm 5$ mCrab
in 4-10 keV energy range for GX 339-4.
In the present paper, we will analyze data of 2006/2007 and 2010 outbursts. 
For spectral analysis, $n_H= 0.5 \times 10^{22} ~ cm^{-2}$ is taken towards this source \citep{M97, K00}. 

(c) Black hole candidate 4U 1543-47 was discovered on August 17, 1971 by Uhuru satellite \citep{M72}. 
Since discovery, X-ray outbursts have been observed
in 1983, 1992, 2002. Optical observation was found by \citet{B83} during 1983 outburst. 
The 2002 outburst commenced on MJD 52442 along with increase in source flux from 0.054 to 1.65 mCrab in 
2-12 keV energy range and was observed simultaneously in X-rays by RXTE 
\citep{M02, P04} in Radio by Molonglo Observatory Synthesis Telescope 
(MOST) \citep{H02, P04} in Optical and Infrared by YALO Telescope \citep{B04}.
%The mass is $2.7$ to $7.5 ~ M_{\odot}$. 
Several multi-wave length observations are reported for 2002 outburst of 4U 1543-47 \citep{K05}.
The blackhole mass $9.4 \pm 0.2 ~ M_{\odot}$, D=7.5 $\pm$ 1.0 kpc, i=20.7 $\pm$ 1.0$^\circ$ was reported by \citet{P04}.
 Fourier resolved spectroscopy is also carried out for the source during this outburst \citep{re06}. 
%For the outburst analysis PCU2 data of RXTE is taken and 1\% systematic error is considered (Park et al. 2004).
For spectral analysis, interstellar absorption is taken to be $n_H = 4.0 \times 10^{21} ~ cm^{-2}$
\citep{D90, P04}. 

(d) XTE J1118+480 was first discovered by All Sky Monitor (ASM) by Remillard et al. (2000) in
March, 2000, while \citet{G00} spectroscopically observed its
optical counter part. This object is observed simultaneously in optical,
and infrared to determine the physical parameters of the black hole \citep{G06}.
The orbital inclination angle is $68^{\circ} \pm 2^{\circ}$. This angle corresponds to a
primary black hole mass of $8.53 \pm 0.60 ~ M_{\odot}$. Distance of the black hole is 
$1.72 \pm 0.10$ kpc. This object has shown outbursts in 2000 \citep{H00, Mc01, C03} and 
2005 \citep{H06, rem05, P05, r05, Z05a, Z05b, Z06}. 
Several authors have discussed disk-jet connection of this compact object \citep{K01, H03b} 
to explain source high energy photons \citep{M01, y05}.
%We took the PCU2 data for the analysis and considered 0.6\% systematic error. 
\citet{rem05} reported this outburst as a faint x-ray outburst as the source flux increases
from 15-19 mCrab in 2-12 keV energy band. Optical \citep{Z05a, Z05b, Z06} and radio \citep{P05, r05} counter 
parts have also been observed. Spectral fitting is done with $n_H = 1.3 \times 10^{20} ~ cm^{-2}$ \citep{BR10}. 

(e) X-Ray transient XTE J1859+226 was discovered by the All Sky Monitor (ASM) onboard 
RXTE on 9 October, 1999 \citep{w99}. \citet{S99} reported the increase of source flux
up to 1.37 mCrab in 2-12 keV energy range.
Mass and inclination angle are $4.5 \pm 0.6 ~ M_{\odot}$ and $70^{\circ}$ respectively
\citep{CS11}. Distance of the compact object is 11 kpc \citep{Z02}.
Spectral analysis is done with $n_H = 2.21 \times 10^{21} ~ cm^{-2}$ \citep{D90}.
%We analyzed the 2000 outburst using the PCU0 and PCU2 data (Casella et al. 2004). 

(f) X-Ray Nova GRO J1655-40 was first discovered by BATSE onboard Compton Gamma Ray Observatory
on 1994 July, 27 \citep{zh94}. Optical counter-part was discovered by \citet{B95}.
Mass and inclination angle are $M=7.02 \pm 0.22 ~ M_{\odot}$, 
$\theta = 69^{\circ}.5 \pm 0^{\circ}.1$ \citep{O97, v98} respectively. 
Distance of the compact object is $3.2 \pm 0.2$ kpc \citep{H95}. In the last week of February, 2005,
this source became active as the source flux increased from 0.3 mCrab to 4.1 mCrab 
in 2-10 keV energy range \citep{M05} and outburst continued for 260 days \citep{s07}. 
Before this, several outbursts were reported of the same source \citep{s99b}. 
Hydrogen column density is fixed at $0.89 \times 10^{22} ~ cm^{-2}$ \citep{zh97}. 
%We analysis the PCU2 data of RXTE PCA.

(g) X-Ray transient XTE J1550-564 was first discovered by ASM in 
1998 September 7 \citep{s98} and by CGRO \citep{w98}. Optical \citep{O98b} and radio \citep{C98}
counterparts are detected shortly after this. Optical photometry revels a binary period of 1.541 $\pm$ 0.009 
days \citep{J01} during the 1998 outburst. During this outburst, presence of a superluminal jet 
is observed along with the massive X-Ray flare \citep{H01}. Later observations revealed
that the black hole has a mass of $10.0 \pm 1.5 ~ M_{\odot}$ and the companion star is a 
late-type subgiant (G8IV-K4III). Binary inclination angle is 
$72^{\circ} \pm 5^{\circ}$ \citep{O02}. Distance of the black hole is 6 kpc \citep{s99a}. 
In 1998, luminosity of the compact object is reported to be 6.8 Crab in 2-12 keV energy range \citep{R98}. 
%For the analysis PCU 1-4 are used (Belloni et al. 2002)
Interstellar absorption $n_H = 0.85 \times 10^{22} ~ cm^{-2}$ is taken \citep{t01a}.

\section{Analysis procedure}

In this paper, we analyzed RXTE data and results are summarized in Table:~1. We chose data sets by MJDs of
outbursts obtained from literature. These data are downloaded from HEASARC, NASA Archive. 
During analysis, we exclude data collected for elevation angles less than $10^{\circ}$, for
offset greater than $0.02^{\circ}$ and those acquired during the South Atlantic Anomaly (SAA) passage. 
%For analysis we have corrected the data for good time and SAA region. 
We selected PCU2 data as it was active most of the time.
Here we calculated $3.0 - 40.0$ keV counts 
in kcts/s and plot them in top panels of Figs. 1-4 described below.
Power Density Spectrum (PDS) is generated by standard FTOOLS task ``powspec'' with a suitable normalization.
Data is binned in $0.01$s to obtain a Nyquist frequency of $50$ Hz as power
beyond this is found to be insignificant. PDSs are normalized to give squared {\it rms} 
fractional variability per Hertz. Evolution of QPO frequency is plotted in second panel 
from top of Figs. 1-4.
  
\subsection{Spectral Analysis}

Spectral analysis of RXTE PCA data is done by using ``standard2" mode data which have $16$ sec time 
resolution. We constrained our energy selection from $3.0$ keV to $40$ keV from the observational data. 
Source spectrum is generated using FTOOLS task ``SAEXTRCT" 
with $16$ sec time bin from ``standard2" data. Background fits file is generated from ``standard2" fits
file by FTOOLS task ``runpcabackest" with standard FILTER file provided with the package.
Background source spectrum is generated using FTOOLS task ``SAEXTRCT" with $16$ sec time bin from
background fits file. Standard FTOOLS task ``pcarsp" is used to generate response file with 
appropriate detector information. Exposure time is corrected for source and background spectra 
with deadtime correction factor. Spectral analysis and modeling was performed using XSPEC (V.12) 
astrophysical fitting package. For model fitting of PCA spectra, we have used a systematic error 
of $0.5\%$. Spectra are fitted with {\it diskbb} and {\it power-law} model along with 
hydrogen column density for absorption. We use {\it Gaussian} for iron line as required for best fit.
During fitting of spectra we adopted a technique introduced by \citet{s99b,P11,P13}, 
to obtain spectral parameters. We calculated 
error-bars at 90\% confidence level in each case. \citet{morri83} reported that interstellar 
absorption mainly affects photons from $0.03$ to $10.0$ keV. It is important to consider this effect in RXTE 
analysis since we need to count injected soft photons up to $10.0$ keV as accurately as possible. 
We chose $n_H$ as appropriate for each candidate to take care of this effect. 

\subsection{Efficiency of Comptonization}

In the literature, usual trend is to plot hardness ratio HR (say, the ratio of 6-40 keV photons divided 
by 2-6 keV photons) when it comes to understand how important Compton scattering is. However,
there are several problems in this definition: (a) HR does not depend on mass of the black hole, 
although hard (Comptonized) and soft photons change their meaning with the mass of the black hole. For instance,
soft photons in lower mass case could be Comptonized photons in higher mass case. 
Thus HR cannot be compared for two objects, or even in two episodes of the same object. 
To circumvent this problem, we define a new parameter \citep{P11,P13} called Comptonization Efficiency (CE) 
which is defined purely on physical ground and independent of any model or mass of the black hole. 
Since for stellar mass black holes we expect peak radiation from standard Keplerian disk to be at 
around $0.5-1$ keV, and slope of power-law tail is defined for $2-10$ keV region, we choose a range 
$0.1-40$ keV to define CE,
both ends being far from the expected values. We compute total number of soft (seed) photons ($N_{BB}$) from 
Keplerian disk from multicolour black body and total number of power-law photons ($N_{PL}$) in ranges 
dynamically determined by best fits obtained after correcting observational data using energy dependent 
absorption due to hydrogen column. Since these ranges are obtained dynamically,  
the ratio CE=$N_{PL}$/$N_{BB}$ is independent of mass of a black hole or any specific model of accretion flow.
Hence CE of two different objects can be compared, which we do in this paper.
Number of black body photons are obtained following \citet{maki86}. Here we count the 
photons within the energy range starting from $0.1$ keV to the best fitted upper energy limit dbb$_e$ 
by first fitting the spectrum with diskbb model alone \citep{s99b}. 
Upper limit of black body energy dbb$_e$ is obtained from the consideration that the  
reduced $\chi^2$ value from resultant fit should be $\sim 1.0$. This upper limit 
may vary with time and from one data set to another. 
 
Comptonized photons $N_{PL}$ are calculated by using the power-law equation given below, 
\begin{equation}
P(E)=N(E)^{-\alpha} ,
\end{equation}
where, $\alpha$ is the power-law index and $N$ is the total photons/s/keV at $1$ keV.
It is reported in \citet{tit94}, that the Comptonization spectrum will have a peak at around
$3 \times T_{in}$, where, $T_{in}$ is  the temperature at the inner edge of the standard disk. 
Thus, the power-law is integrated from $3 \times T_{in}$ to $40$ keV to
calculate total rate of Comptonized photons (photons/s). $T_{in}$ and $\alpha$ comes from spectral fit
of each data.  Here we assume that the black hole is non-rotating. For rotating flows the 
corresponding spectral properties of the standard disk has to be used.
 
%We have verified our fitting procedure with that of the {\it  simpl} model. The `fracsctr' value in that 
%model is obtained around 10-40\%, since simpl models computes the photons from $2.0$ keV. If we 
%also count soft photons from $2$ KeV, the value of CE matches with the value of 'fracsctr'. 
%However, in order to compute soft photons accurately, we needed to  compute them
%from $0.1$ KeV, much below the limit of RXTE. Our computed CE purely from observation is in agreement with 
%the findings of Two Component Advective FLow (TCAF) solution (CT95) that the interception of the soft photon 
%is at the most a few percent.  
%The presence of line emissions are taken care of by the usual Gaussian fits. 

\begin{table}
%\scriptsize{
\addtolength{\tabcolsep}{6.0pt}
 \label{symb}
 \begin{tabular}{@{}cccc}
%\multicolumn{12}{|c|}
%{Table 1: Data sets analyzed for all the compact objects.}\\
\hline 
Object & Outburst & Analysis Time & Obs ID \\ 
\hline
&  &  & 30188-06-03-00, 30188-06-01-00 30188-06-01-02 30188-06-04-00   \\ 
&  & 08/09/1998 & 30188-06-05-00, 30188-06-07-00, 30188-06-09-00, 30188-06-11-00   \\ 
XTE J1550-564& 1998 & to  & 30191-01-01-00, 30191-01-02-00, 30191-01-05-00,  30191-01-06-00    \\ 
&  & 10/10/1998 & 30191-01-08-00, 30191-01-09-00, 30191-01-10-00, 30191-01-13-00,   \\ 
&  &  & 30191-01-16-00, 30191-01-18-01, 30191-01-27-00   \\ 
\hline
&  &  &  50137-02-01-00, 50137-02-02-00, 50137-02-03-00, 50137-02-04-00,  \\ 
&  & 10/04/2000 &  50137-02-05-00, 50137-02-06-00, 50134-02-01-00, 50134-02-01-01,  \\ 
XTE J1550-564 & 2000 & to &  50134-02-02-00, 50134-02-03-00, 50134-02-03-01, 50134-02-04-00,   \\ 
&  & 26/04/2000 &  50134-02-05-00, 50134-02-07-01, 50134-01-01-00, 50134-01-03-00,    \\ 
&  &  &  50134-01-05-00, 50135-01-01-00, 50135-01-02-00, 50135-01-04-00,   \\ 
&  &  &  50135-01-05-00, 50135-01-07-00, 50135-01-08-00  \\ 
\hline
&  &  & 92052-07-04-00, 92052-07-05-00, 92428-01-01-00, 92052-07-06-00,   \\ 
&  &  & 92052-07-06-01, 92428-01-02-00, 92428-01-03-00, 92035-01-01-02,    \\ 
&  &  & 92035-01-02-00, 92035-01-02-08, 92035-01-03-01, 92428-01-04-04,    \\ 
&  &  28/12/2006 & 92035-01-04-02, 92085-01-01-06, 92085-01-02-03, 92085-01-02-06,    \\ 
GX 339-4 & 2007 & to & 92085-01-03-02,  92085-01-04-13, 92085-01-04-02, 92085-02-01-01,    \\ 
&  & 26/05/2007 & 92085-02-01-04,  92085-02-02-00, 92085-02-02-01, 92085-02-03-00,    \\ 
&  &  & 92085-02-03-01,  92085-02-04-00,  92085-02-04-02, 92085-02-05-01,    \\ 
&  &  & 92085-02-05-02, 92704-03-02-00, 92704-03-03-00, 92704-03-05-01,   \\ 
&  &  & 92704-03-07-01, 92704-03-09-02, 92704-03-11-01 \\ 
&  &  & 92704-03-13-02,  92704-03-15-00 \\ 
\hline
&  &  & 95409-01-02-01, 95409-01-07-01, 95409-01-10-03, 95409-01-13-04,   \\ 
GX 339-4 & 2010 & 18/01/2010 & 95409-01-14-04, 95409-01-15-01, 95409-01-15-02, 95409-01-15-04 \\ 
 &  &  to & 95409-01-15-06, 95409-01-16-00, 95409-01-16-03, 95409-01-17-01,  \\ 
&  & 08/05/2010 & 95409-01-17-03, 95409-01-17-05, 95409-01-18-00, 95335-01-01-00  \\ 
\hline
H 1743-322 & 2008 & 16/01/2008 to & 93427-01-01-00, 93427-01-02-01, 93427-01-02-02, 93427-01-03-00,  \\ 
&  & 04/02/2008 & 93427-01-03-01, 93427-01-03-04, 93427-01-04-01, 93427-01-04-02 \\ 
\hline
&  &  29/05/2009 & 94413-01-02-00, 94413-01-02-02, 94413-01-02-05, 94413-01-02-03  \\ 
H 1743-322 & 2009 & to & 94413-01-03-00, 94413-01-03-01  \\ 
&  & 07/07/2009 & 94413-01-03-04, 94413-01-04-00, 94413-01-04-04, 94413-01-07-01  \\ 
\hline
&  &  & 40124-01-05-00, 40124-01-09-00, 40124-01-12-00, 40124-01-17-00, \\ 
&  &  & 40124-01-20-00, 40124-01-23-01, 40124-01-29-00, 40124-01-33-01,  \\ 
&  & 12/10/1999 & 40124-01-38-01, 40124-01-44-00, 40122-01-04-00, 40124-01-54-00,  \\ 
XTE 1859+226 & 2000 & to & 40124-01-55-03, 40124-01-57-01, \\ 
&  & 26/02/2000 & 40124-01-58-01, 40124-01-59-00, 40124-01-60-00, 40124-01-61-00,   \\ 
&  &  & 40124-01-62-00, 40124-01-63-01, 40124-01-64-01, 40124-01-65-00,   \\ 
&  &  & 40124-01-66-00, 40440-01-01-00, 40440-01-02-00, 40440-01-03-00  \\ 
\hline
&  &  13/01/2005 & 90111-01-01-00, 90011-01-01-00, 90011-01-01-02, 90011-01-01-09,  \\ 
XTE 1118+480 & 2005 & to & 90011-01-01-11, 90011-01-01-06,  \\ 
&  & 26/01/2005 & 90111-01-02-03, 90111-01-02-07, 90111-01-02-10, 90111-01-02-11  \\ 
\hline
&  &  & 91404-01-01-01, 91404-01-01-04, 91702-01-01-03, 91702-01-01-05    \\ 
&  &  & 90704-04-01-01, 90704-04-01-00, 91702-01-02-00, 91702-01-02-01  \\ 
&  &  & 91702-01-02-03, 91702-01-13-00, 91702-01-19-00, 91702-01-29-00  \\ 
&  &  & 91702-01-36-01, 91702-01-44-03, 91702-01-44-01, 91702-01-44-04  \\ 
&  & 06/03/2005 & 91702-01-52-02,  91702-01-56-03, 91702-01-60-00, 91702-01-63-00  \\ 
GRO J1655-40 & 2005 & to & 91702-01-68-01, 91702-01-70-00, 91702-01-72-02, 91702-01-77-00  \\ 
&  & 19/09/2005 & 91702-01-78-01, 91702-01-83-01, 91702-01-85-00, 91702-01-90-02  \\ 
&  &  & 91702-01-94-00, 91702-01-95-02, 91702-01-01-10, 91702-01-05-11  \\ 
&  &  & 91702-01-16-10,  91702-01-24-10, 91702-01-29-10, 91702-01-33-11  \\ 
&  &  & 91702-01-41-11, 91702-01-50-10, 91702-01-51-13, 91702-01-74-00  \\ 
&  &  & 91702-01-76-00, 91702-01-79-00, 91702-01-80-00, 91702-01-80-01  \\ 
\hline
&  & 17/06/2002 & 70133-01-01-00, 70133-01-02-00, 70133-01-04-00, 70132-01-01-00  \\ 
4U1543-47 & 2002 & to & 70133-01-12-00, 70133-01-07-00, 70133-01-10-00  \\ 
&  & 25/07/2002 & 70133-01-15-00, 70133-01-18-00, 70133-01-20-00, 70133-01-26-00   \\ 
&  &  & 70133-01-28-00, 70133-01-31-00, 70124-02-03-00  \\ 
\hline
%-& - &-  & --   \\ 
\end{tabular}
%}
\caption{Observation IDs analyzed in this paper.}
\end{table}

%%%%%%%%%%%%%%%%%%%Fig.1%%%%%%%%%%%%%%%%%%%%%%%%%
\begin{figure}
\includegraphics[width=6.35truecm,angle=270]{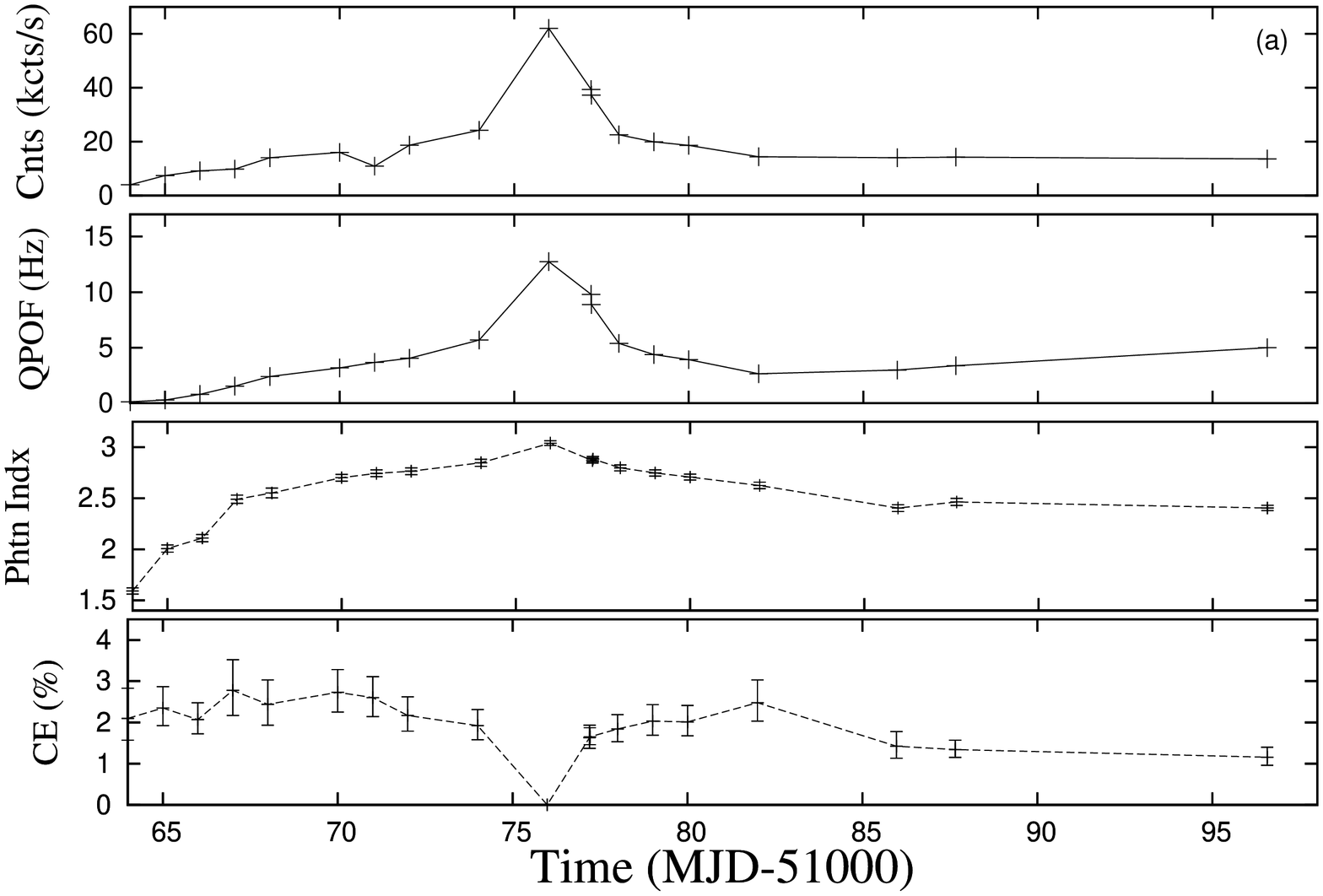}
\includegraphics[width=6.35truecm,angle=270]{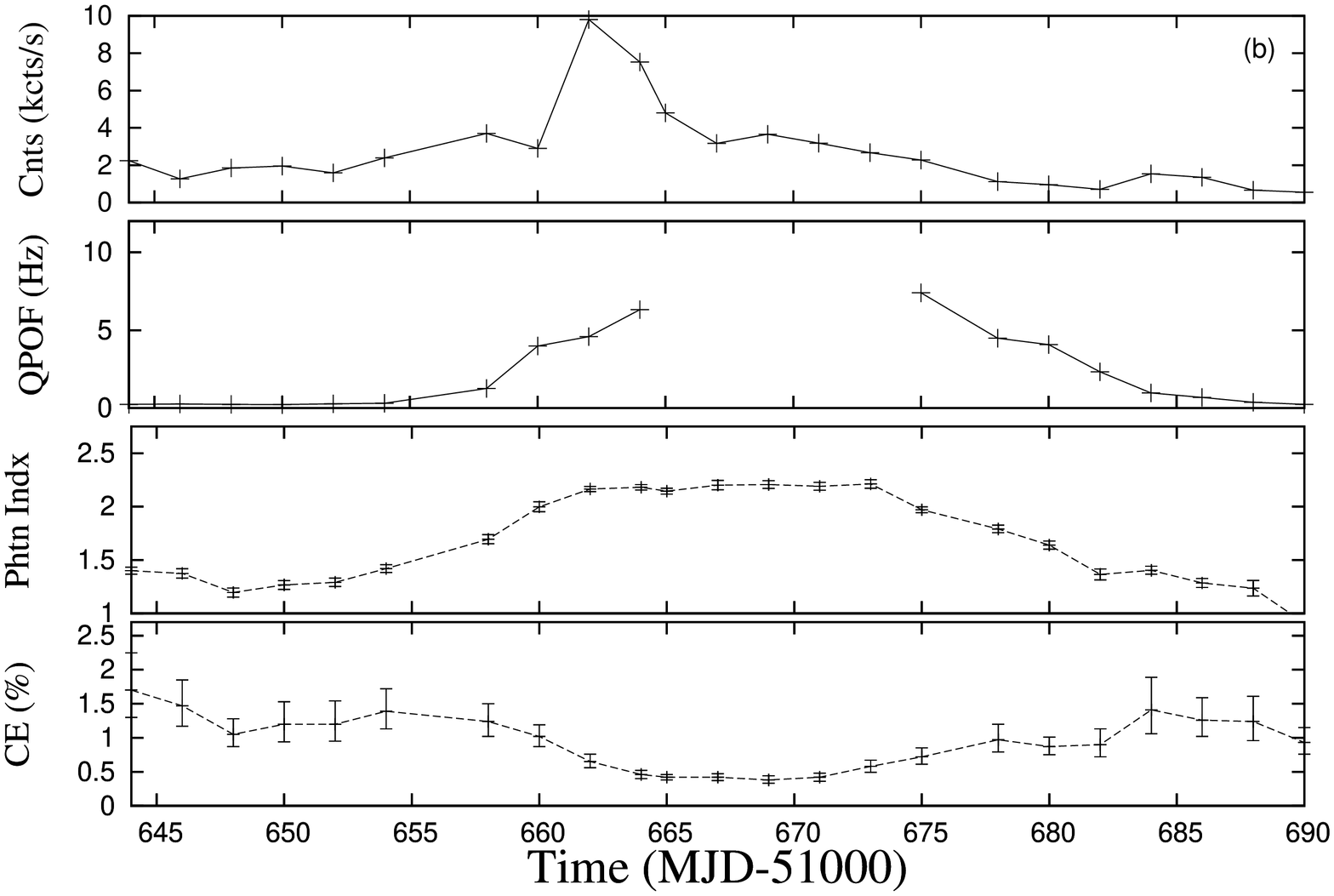}
\caption{Simultaneous variation of $2.0 - 40.0$ keV counts in kcts/s (upper panel), QPO frequency in Hz (second panel), 
spectral slope (third panel) and the CE (lower panel) of 
(a) XTE J1550-564 during its 1998 outburst and (b) XTE J1550-564 during its 2000 outburst.}
\label{fig1550}
\end{figure} 

\begin{figure}
\includegraphics[width=6.35truecm,angle=270]{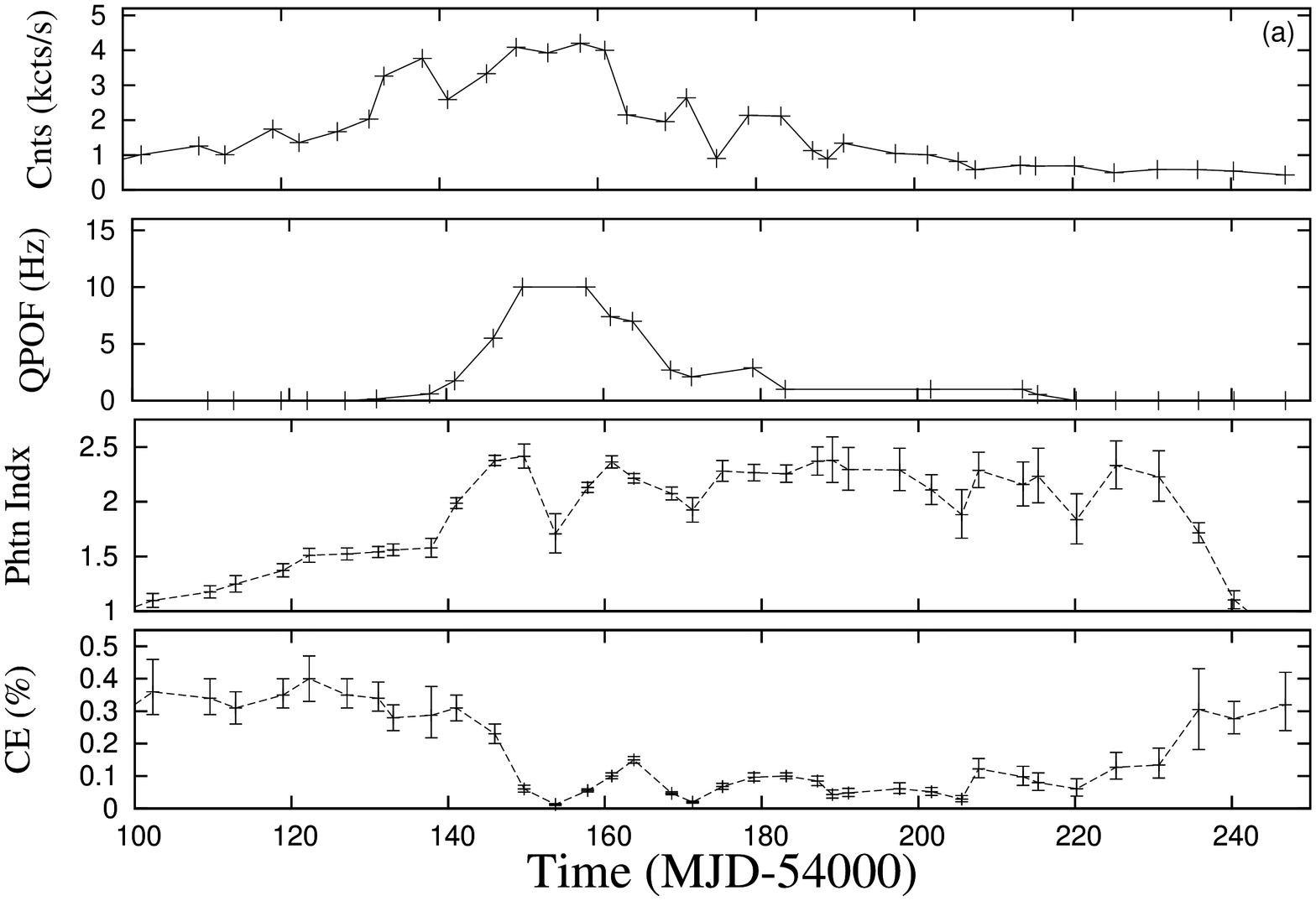}
\includegraphics[width=6.35truecm,angle=270]{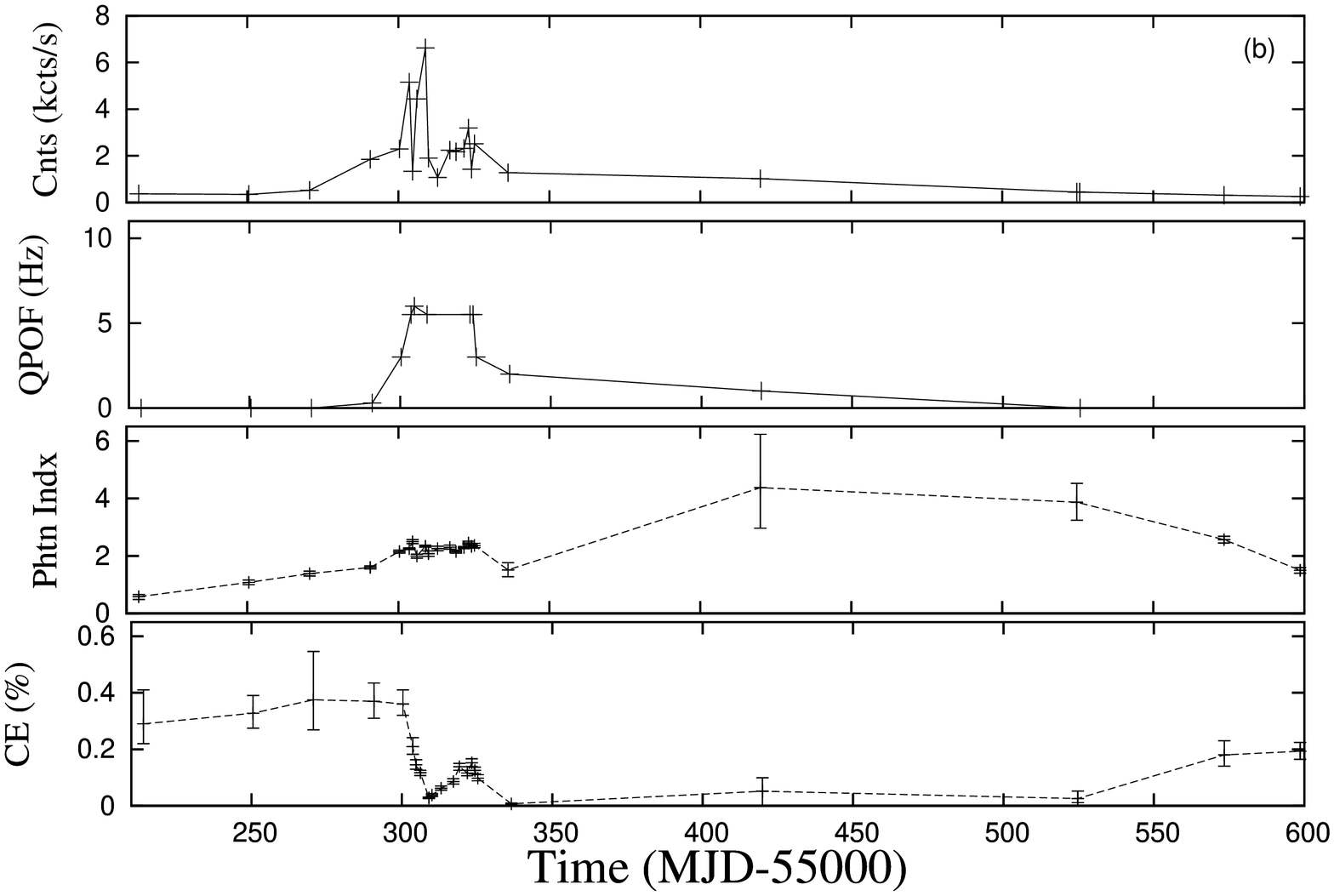}
\caption{Same as in Fig. 1, except
(a) GX 339-4 during its 2007 outburst and (b) GX 339-4 during its 2010 outburst.}
\label{figgx}
\end{figure} 

\begin{figure}
\includegraphics[width=6.35truecm,angle=270]{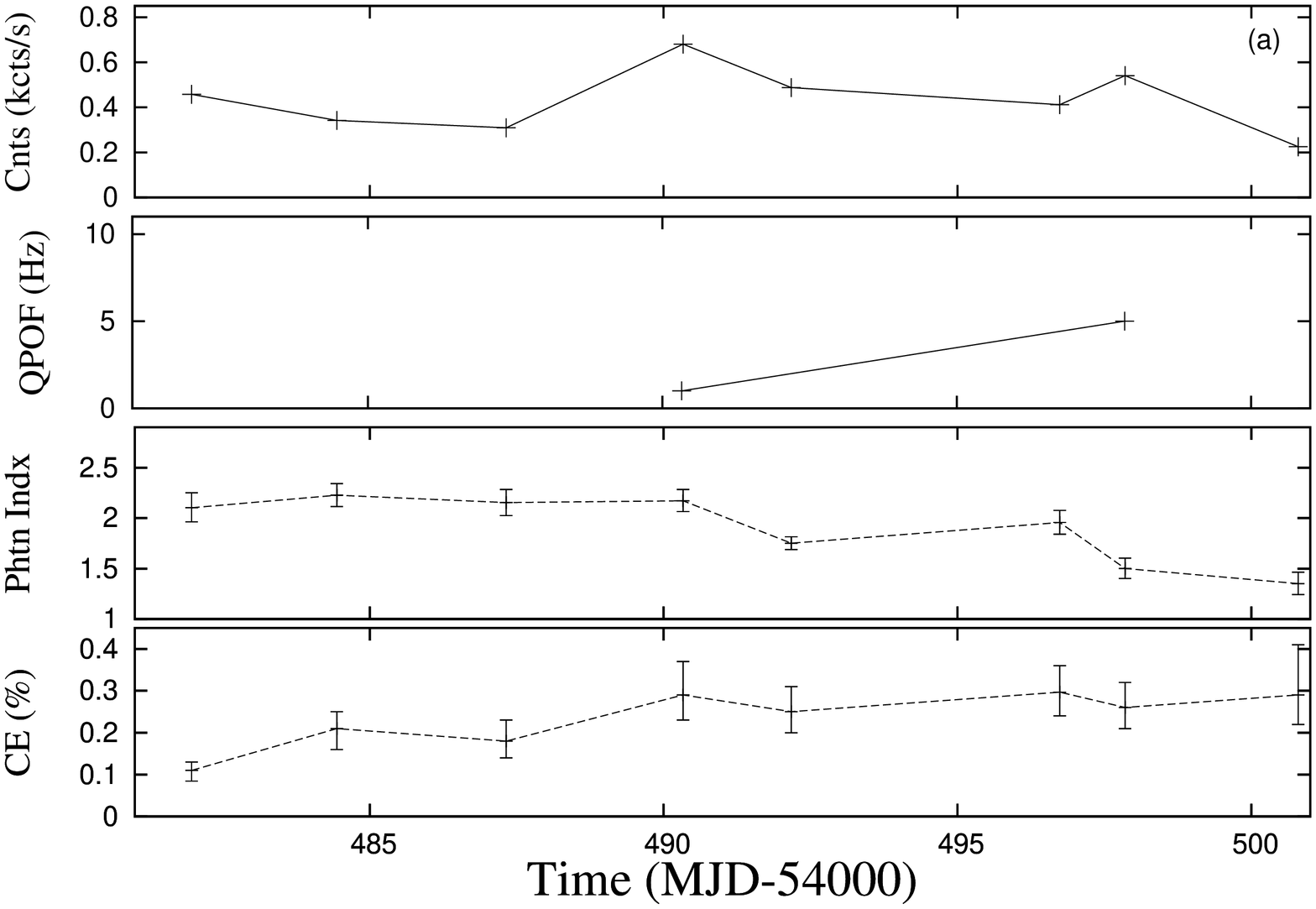}
\includegraphics[width=6.35truecm,angle=270]{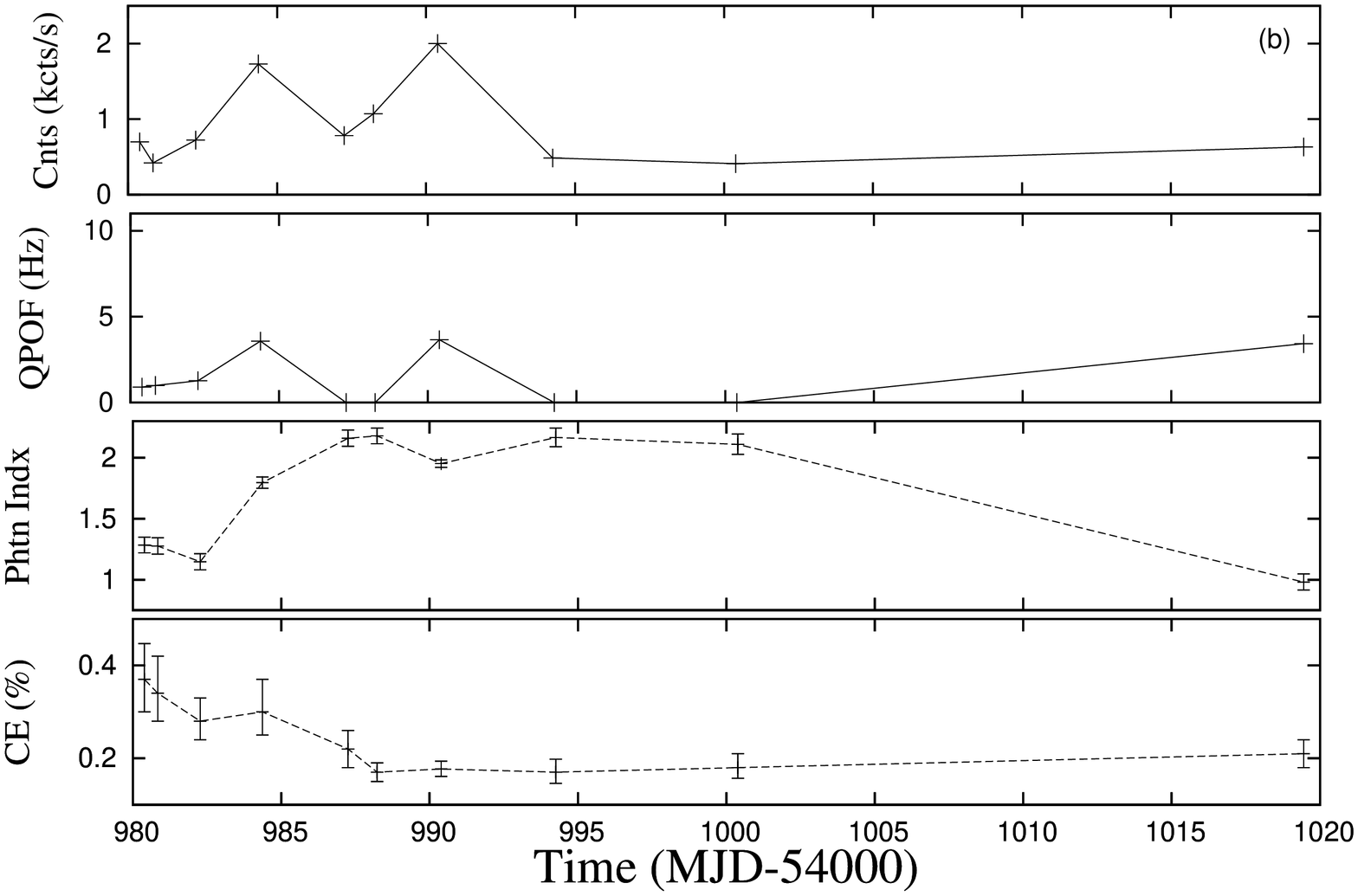}
\caption{Same as in Fig. 1, except
(a) H 1743-322 during its 2008 outburst (b) H 1743-322 during its 2009 outburst.}
\label{figh}
\end{figure} 

\begin{figure}
\includegraphics[width=6.35truecm,angle=270]{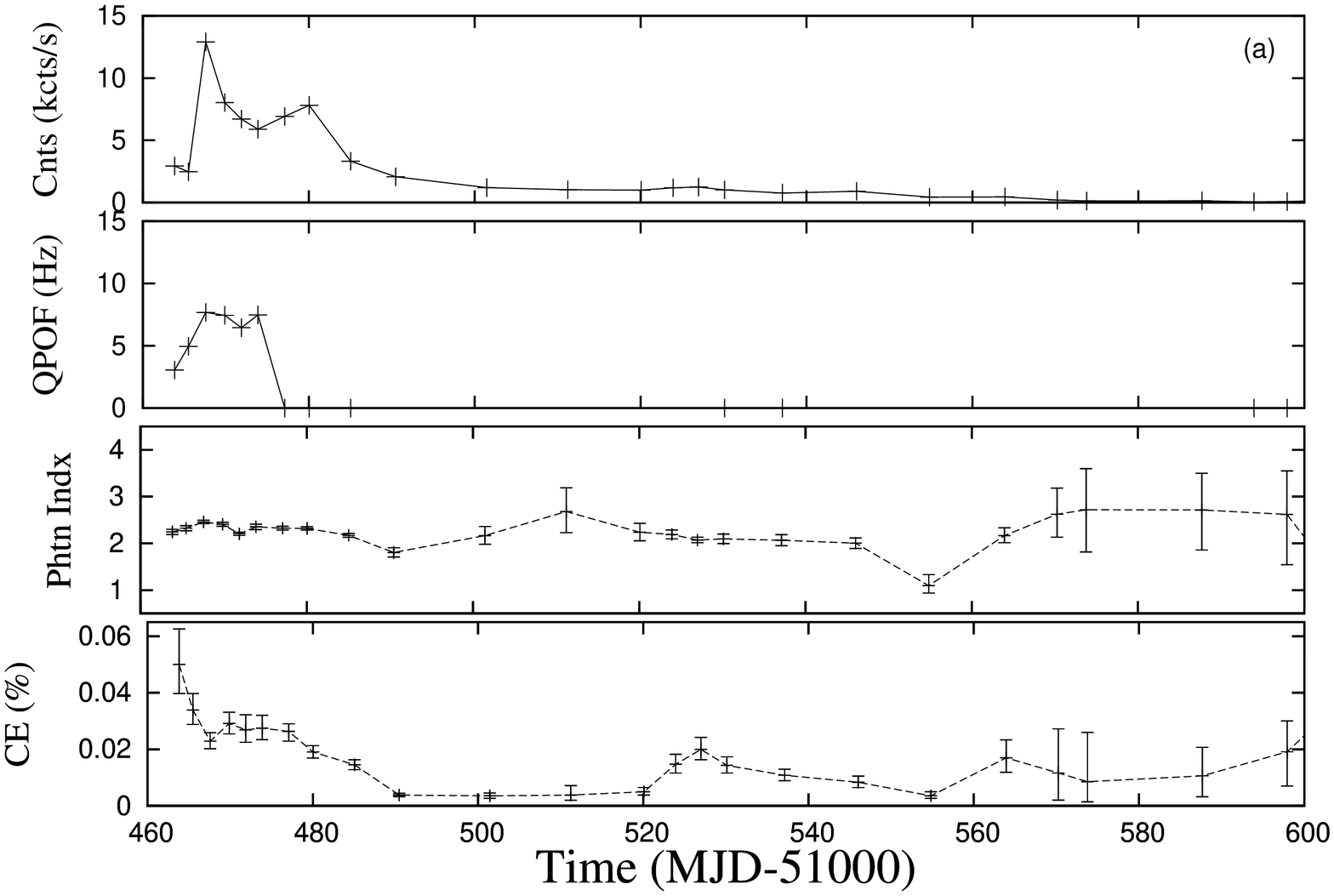}
\includegraphics[width=6.35truecm,angle=270]{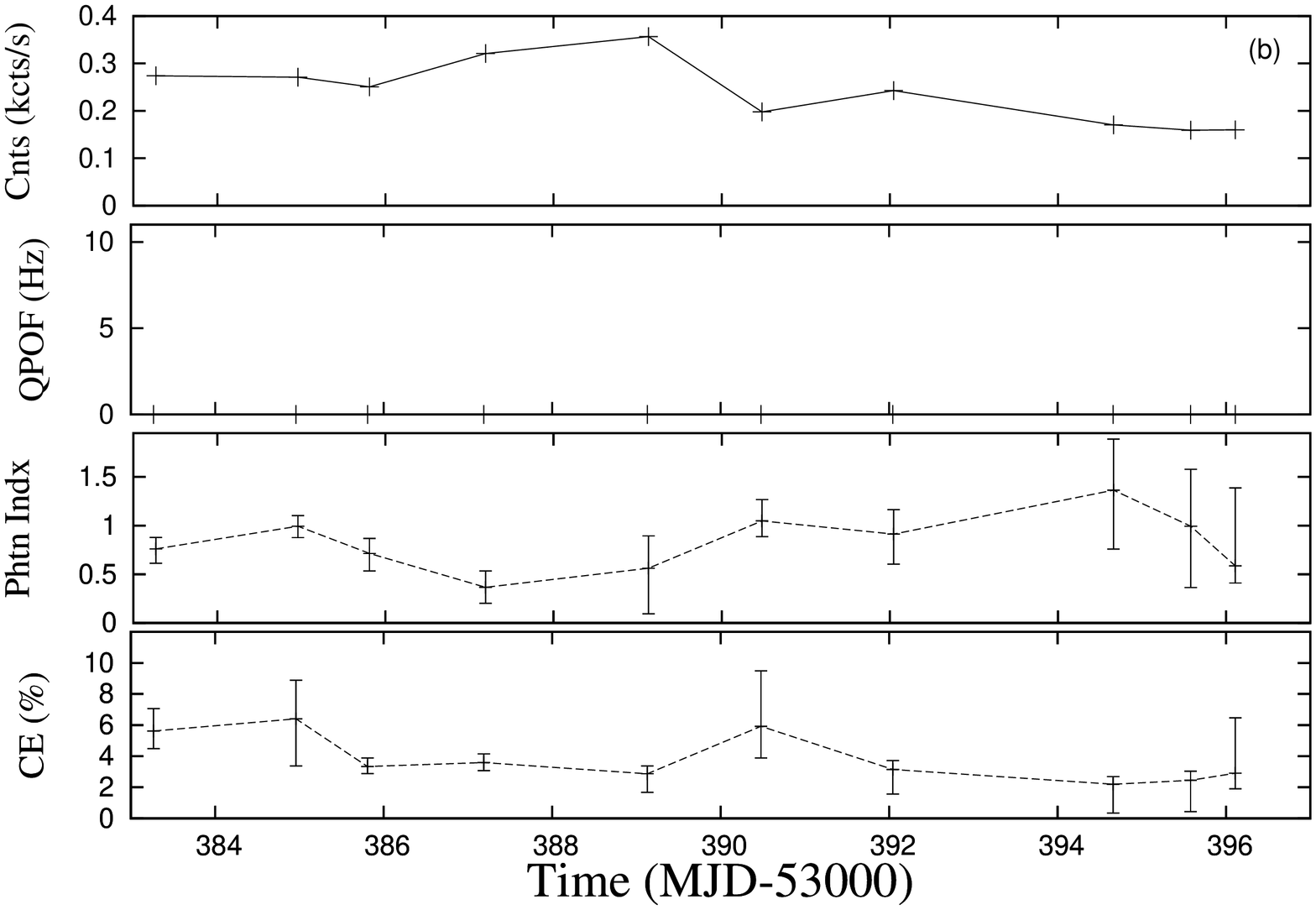}
\includegraphics[width=6.35truecm,angle=270]{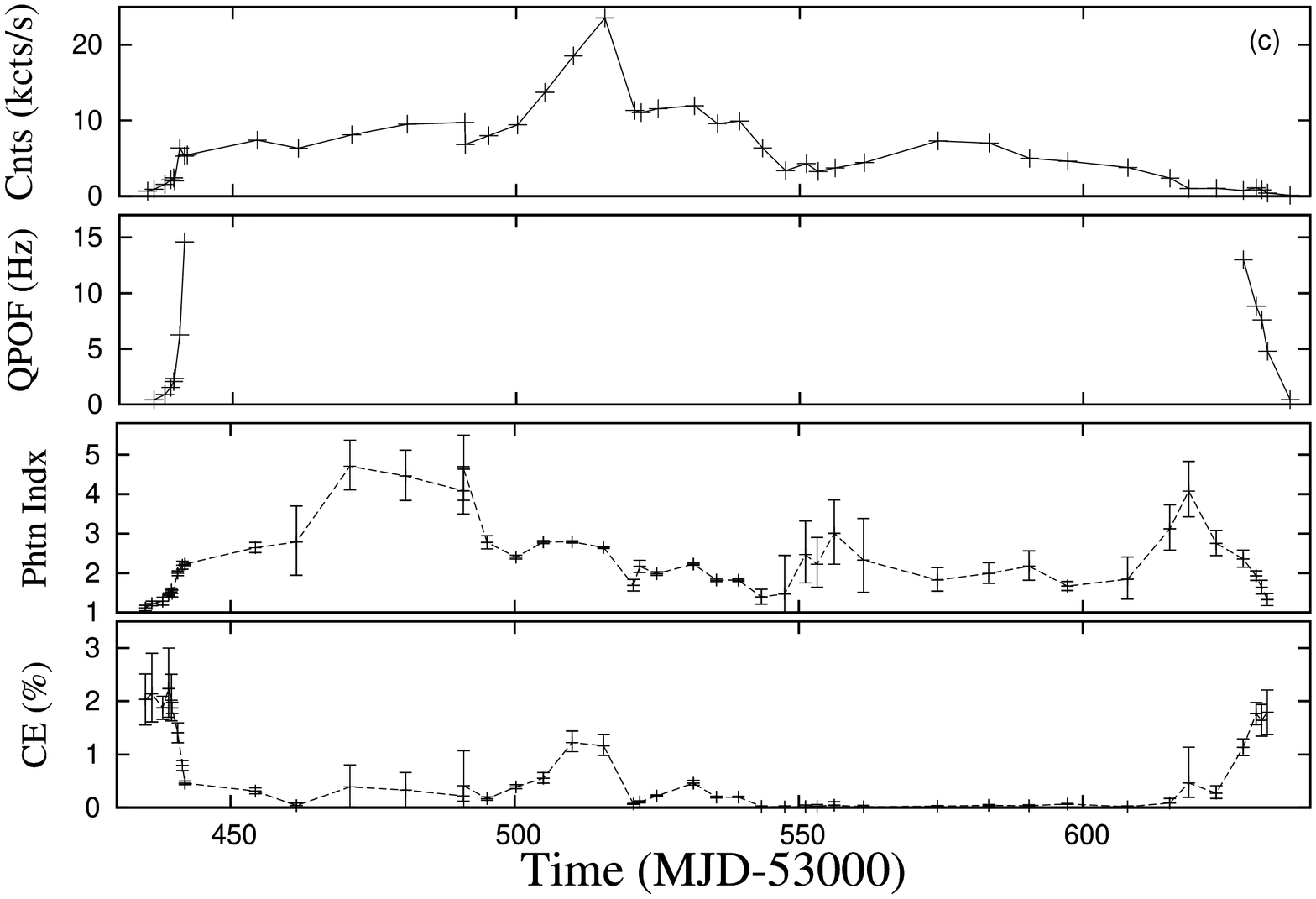}
\includegraphics[width=6.35truecm,angle=270]{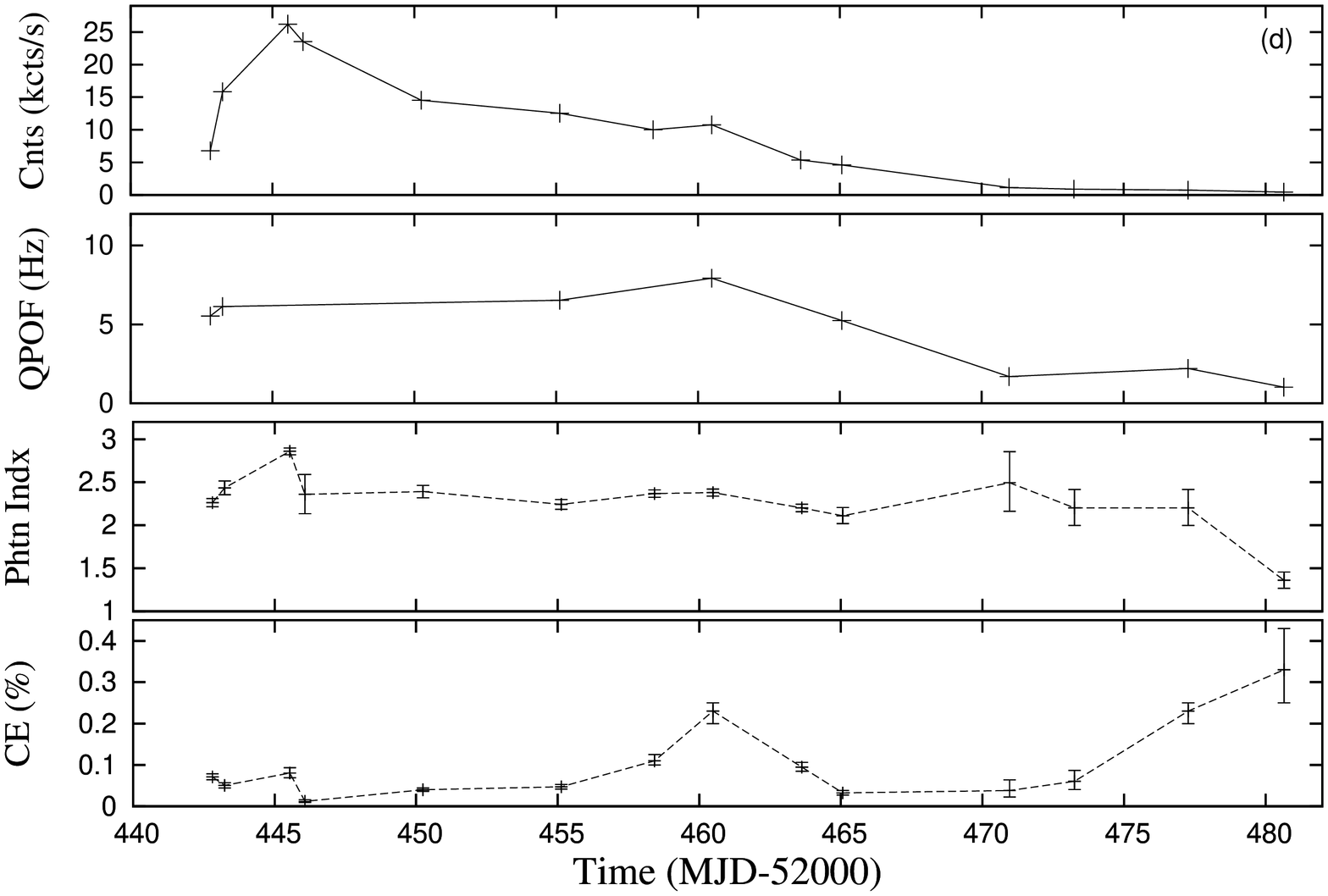}
\caption{Same as in Fig. 1, except
(a) XTE 1859+226 during its 2000 outburst, (b) XTE 1118+480 during its 2005 outburst, 
(c) GRO J1655-40 during its 2005 outburst and (d) 4U 1543-47 during its 2002 outburst.}
\label{figxte}
\end{figure} 

\section{Results \& discussion}

Observation IDs of data sets analyzed in this paper are given in Table:~1. 
In Table:~2, spectral fitting parameters of sample datasets are shown. 
%In the first column, the name 
%of the Compact Object (outburst year) is given. In the second column,
%MJDs of the dataset parameters are given in the corresponding row. The third column shows the QPO frequency in Hz.
%The fourth column shows the black body temperature in keV. 
%The fifth column shows the upper energy range ($dbb_e$) up to which 
%the blackbody photons are calculated. The sixth column shows the $\tilde \chi^2 (dof)$ for the disk blackbody
%parameters. The seventh column shows the number of soft photons calculated from the disk blackbody parameters
%within $0.1$ - $dbb_e$ keV. Eighth column shows the power-law index. The ninth column shows the $\tilde \chi^2$
%for the fitting. The tenth column shows the calculated number of photons within $3 \times T_{in}$ - $40.0$ keV.
%The eleventh column shows the CE. 
Error-bars are calculated at 90\% confidence level in each case. To ease discussion, we keep two component
advective flow (TCAF) model of \citet{C95} in the back of our mind. Of course, our result does not depend on 
any specific model, as long as there are sources of seed photons and hot electrons for inverse 
Comptonization of these seed photons.

\begin{table}
%\scriptsize{
\addtolength{\tabcolsep}{-1.95pt}
 \label{param}
\begin{tabular}{@{}|c|c|c|c|c|c|c|c|c|c|c|}
%\multicolumn{12}{|c|}
%{Table 1: Data sets analyzed for all the compact objects.}\\
\hline 
Compact & Time & QPO & $T_{in}$ &  $dbb_e$ & $\tilde \chi^2$ & Soft & Power-law & $\tilde \chi^2$ & Hard & CE\\ 
Object  &  & &  &  &       & Phtns & index & & Phtns &  \\ 
(Outburst) &MJD& Hz & (keV) &  (keV) & (dofs) & kphtns/s& & (dofs) & (kphtns/s) & (\%)\\ 
\hline
XTE J1550-564(1998) & 51065 & 0.3 & $1.74^{+0.10}_{-0.09}$ & %$10.39^{+1.17}_{-1.10}$ & 
4.75 & 1.5(7) & $43.89^{+8.15}_{-6.46}$ & $2.00^{+0.03}_{-0.03}$ & 1.3(71) & $1.03^{+0.04}_{-0.04}$ & $2.35^{+0.52}_{-0.43}$ \\
\hline
%XTE J1550-564(1998) & 51076 & 12.7 & $0.99^{+0.05}_{-0.04}$ &$38.87^{+4.35}_{-3.99}$ & 8.0 & 0.97(10) & $108.21^{+16.04}_{-13.36}$ & $3.04^{+0.02}_{-0.02}$ & 1.1(79) & $0.001^{+0.0014}_{-0.0008}$ & $0.001^{+0.0003}_{-0.0005}$ \\
%\hline
XTE J1550-564(2000) & 51648 & 0.2 & $1.60^{+0.14}_{-0.12}$ & %$12.35^{+1.27}_{-1.25}$ & 
6.0 & 1.7(8) & $5.27^{+0.10}_{-0.08}$ & $1.19^{+0.04}_{-0.04}$ & 1.5(74) & $0.06^{+0.001}_{-0.001}$ & $1.05^{+0.23}_{-0.18}$ \\
\hline
XTE J1550-564(2000) & 51665 & - & $1.02^{+0.02}_{-0.02}$ & %$40.45^{+2.77}_{-2.62}$ & 
8.5 & 1.4(9) & $130.71^{+9.9}_{-10.0}$ & $2.14^{+0.03}_{-0.03}$ & 0.9(69) & $0.55^{+0.02}_{-0.01}$ & $0.42^{+0.04}_{-0.04}$ \\
\hline
GX 339-4(2007) &54141& 1.7 & $1.85^{+0.06}_{-0.05}$ & %$12.87^{+0.17}_{-0.17}$ & 
5.75 & 1.36(8) & $49.38^{+4.6}_{-4.1}$& $1.99^{+0.05}_{-0.05}$ & 1.6(74) & $0.15^{+0.006}_{-0.005}$ & $0.31^{+0.04}_{-0.04}$\\
\hline
GX 339-4(2007) &54158& 10.0 & $0.91^{+0.01}_{-0.01}$ & %$29.42^{+0.97}_{-0.95}$ & 
8.5 & 1.8(10) & $574.41^{+21.10}_{-20.14}$& $2.13^{+0.05}_{-0.04}$ & 0.9(74) & $0.32^{+0.02}_{-0.01}$ & $0.06^{+0.004}_{-0.003}$\\
\hline
GX 339-4(2010) &55291& 0.3& $1.54^{+0.11}_{-0.10}$ & %$11.59^{+0.14}_{-0.05}$ & 
6.0 & 0.98(9) & $24.49^{+3.86}_{-3.15}$& $1.60^{+0.05}_{-0.05}$& 1.2(79) &$0.09^{+0.003}_{-0.003}$ & $0.37^{+0.06}_{-0.06}$\\   
\hline
GX 339-4(2010) &55313& - & $0.93^{+0.01}_{-0.01}$ & %$16.19^{+0.56}_{-0.54}$ & 
8.5 & 0.98(10) & $186.29^{+7.53}_{-7.15}$ & $2.26^{+0.08}_{-0.08}$ & 1.1(79)& $0.11^{+0.01}_{-0.009}$ & $0.06^{+0.08}_{-0.07}$ \\   
\hline
H 1743-322(2008) & 54498 & 5.0 & $1.61^{+0.08}_{-0.07}$ & %$2.40^{+0.26}_{-0.24}$ & 
5.75 & 1.2(8) & $11.12^{+1.8}_{-1.5}$ & $1.5^{+0.1}_{-0.1}$ & 0.99(79) & $0.03^{+0.002}_{-0.001}$ & $0.26^{+0.06}_{-0.05}$ \\
\hline
H 1743-322(2009) & 54990 & 3.65 & $1.18^{+0.03}_{-0.02}$ & %$13.54^{+0.77}_{-0.74}$ & 
5.25 & 0.7(9) & $136.00^{+9.69}_{-8.84}$ & $1.95^{+0.03}_{-0.03}$ &1.5(79) & $0.24^{+0.006}_{-0.006}$ &  $0.18^{+0.02}_{-0.02}$ \\
\hline
XTE J1859+226(2000) & 51464 & 3.0 & $1.32^{+0.08}_{-0.07}$ & %$19.15^{+2.56}_{-2.35}$ & 
5.5 & 1.1(9) & $668.59^{+130.78}_{-103.10}$ & $2.24^{+0.06}_{-0.06}$ & 1.6(79) & $0.33^{+0.02}_{-0.02}$ & $0.05^{+0.01}_{-0.01}$\\
\hline
XTE J1859+226(2000) & 51480 & - & $1.11^{+0.03}_{-0.03}$ & %$60.18^{+4.05}_{-3.84}$ & 
7.5 & 0.8(8) & $3832.64^{+320.66}_{-288.11}$ & $2.32^{+0.03}_{-0.03}$ & 1.1(75) & $0.73^{+0.03}_{-0.02}$ & $0.02^{+0.002}_{-0.002}$\\
\hline
XTE J1118+480(2005) & 53395 & - & $1.05^{+0.05}_{-0.04}$ & %$10.30^{+1.34}_{-1.20}$ & 
6.5 & 0.94(9) & $312.01^{+47.95}_{-39.70}$ & $2.75^{+0.03}_{-0.03}$ & 1.6(75) & $6.33^{+0.28}_{-0.26}$ & $2.20^{+0.40}_{-0.34}$\\
\hline
GRO J1655-40(2005) & 53439 & 1.5 & $1.54^{+0.17}_{-0.13}$ & %$2.46^{+0.47}_{-0.41}$ & 
5.75 & 0.9(7) & $4.36^{+1.37}_{-0.95}$ & $1.46^{+0.03}_{-0.03}$ & 1.3(75) & $0.10^{+0.002}_{-0.002}$ & $2.24^{+0.7}_{-0.5}$ \\
\hline
GRO J1655-40(2005) & 53591 & - & $1.05^{+0.003}_{-0.003}$ & % $22.48^{+0.25}_{-0.25}$ & 
10.0 & 0.8(11) & $111.56^{+1.19}_{-1.17}$ & $2.17^{+0.39}_{-0.35}$ & 1.2(79) & $0.04^{+0.02}_{-0.02}$ & $0.03^{+0.02}_{-0.02}$ \\
\hline
4U 1543-47(2005) & 52460 & 7.9 & $1.03^{+0.02}_{-0.02}$ & %$35.07^{+2.34}_{-2.31}$ & 
5.5 & 1.6(7) & $401.31^{+30.17}_{-27.43}$ & $2.38^{+0.04}_{-0.04}$ & 1.3(79) & $0.91^{+0.05}_{-0.04}$ & $0.23^{+0.02}_{-0.03}$\\
\hline
%GRS 1915+105(-) &  & &  & & &  &  & &   &\\
%\hline
\end{tabular}
%}
\caption{Parameters for spectral fits of sample dataset with diskbb plus power-law 
models. $T_{in}$ is black body temperature obtained from fitting.
dbb$_e$ is  upper limit energy of disk blackbody spectrum. 
`Soft' column gives blackbody photons in $0.1-dbb_e$ keV. The column `power-law'
contains power-law index $\alpha$ obtained from our fitting. Column `hard photons' contains
rate at which Comptonized photons are emitted in the range $3 \times T_{in}$ to $40$ keV. 
CE is Comptonizing efficiency.}
\end{table}

\subsection{XTE J1550-564}

Outbursts of 1998 and 2000 of Galactic black hole XTE J1550-564 are analyzed in this paper.
1998 outburst is analyzed from MJD 51064 (08/09/1998) to MJD 51097 (10/10/1998). 
Result is shown in Fig.~\ref{fig1550}a. \citet{C09} reported that 
QPO was always observed and thus the oscillating shock in so-called `propagating' shock model \citep{DD10}
does not disappear behind the horizon or the shock does not become weak enough. 
The post-shock region is known as the CENtrifugal pressure supported BOundary Layer or the CENBOL.
It is the Compton cloud in the two component advective flow (TCAF) model \citep{C95}.
In initial stages of the outburst, CE was around $2-3$\%. CE gradually dropped 
to $\sim 0.001$\% as peak of outburst is reached and photon index became highest. 
Physically, this means a gradual decrease of size of CENBOL as Keplerian disk rate rises
and cools CENBOL down. After the peak, CE started to increase to $2-3$\% in the decline phase, 
though it finally converged to $\sim 1$\%. Variation of QPOs during this outburst is discussed in details in \citet{C09}. 
    
We then analyzed 2000 outburst of the same source from MJD 51644 (10/04/2000) to MJD 51690 (26/05/2000). 
Result is shown in Fig.~\ref{fig1550}b. Light curve looks qualitatively the same as that of the 1998 outburst.
During this outburst, CE varies initially between $1.0-1.5$\% but after MJD 51660, CE 
is reduced to less than $0.5$\%. This indicates that oscillating shock was still present. 
After a few days, CE again increased to $\sim 1.5$\% before settling to a $\sim 1$\%. As the shock 
recedes from the black hole, optical depth initially rises, but then goes down as CENBOL density
drops rapidly. This may be the cause for $CE$ to rise first and then to come down at $\sim 1$\% 
in both outbursts.

\subsection{GX 339-4}
We analyzed 2007 and 2010 outbursts of GX 339-4. 2007 outburst is analyzed 
from MJD 54097 (28/12/2006) to MJD 54247 (26/05/2007). Result is shown in Fig.~\ref{figgx}a. 
During initial stage of rising phase, CE remained roughly constant at $\sim 0.3-0.4\% $ for about $40$ days.
Constancy of CE may mean that while CENBOL is getting smaller, density is becoming larger
and thus optical depth remains roughly constant to intercept roughly similar number of soft photons.
After that, CE decreased sharply when the CENOBOL disappeared. A sporadic shock is formed during soft 
and hard-intermediate states, raising CE. QPO remained absent for the next 80 days when CE varied between 
$\sim 0.05-0.1$\%. After that, CE is increased to $\sim 0.3$\%, close to its initial value.

2010 outburst is analyzed from MJD 55214 (18/01/2010) to MJD 55324 (08/05/2010). Result is shown in 
Fig.~\ref{figgx}b. During initial phase, CE remained constant at around $0.3$\% for 80 days. After 
that CE decreases sharply below $0.1$\%. 

\subsection{H 1743-322}
We analyzed outbursts of 2008 and 2009 of H 1743-322.
2008 outburst is analyzed from MJD 54482 (16/01/2008) to MJD 54500 (04/02/2008). 
Only declining phase is observed by RXTE for this outburst.
Result is shown in Fig.~\ref{figh}a. Observation started when the object is in a softer state having a 
high photon index. CE also started from a minimum value of $\sim 0.1$\%. Eventually,
at the end of the outburst, CE increased to the high state value of $\sim 0.3$\%. 
    
2009 outburst is analyzed from MJD 54980 (29/05/2009) to MJD 55019 (07/07/2009). 
During rising phase, CE decreased from $\sim 0.4$\% to $0.15$\% within $10$ days. 
During declining phase, CE rises very slowly $0.2$\%. Result is shown in Fig.~\ref{figh}b. In this case
also, near constancy of CE indicates that optical depth is nearly constant even through
shock wave is receding as is obvious from time variation of QPOs \citep{C08, C09, n12}.

\subsection{XTE 1859+226}
We analyzed 2000 outburst of XTE 1859+226. Compact object is
analyzed from MJD 51463 (12/10/1999) to MJD 51600 (26/02/2000). Result of analysis is 
shown in Fig.~\ref{figxte}a. During onset phase, CE dropped sharply 
from $\sim 0.06$\% to $0.003$\% within 30 days. After that, CE remained low for around 20 days. 
There is a bumps in CE at around MJD 53510, and they could be due to higher accretion rates causing the 
CENBOL to swell by radiation pressure. In the last phase, CE returned back to pre-burst value 
corresponding to a hard state. During this outburst, QPO frequency increased gradually 
from 3 Hz to 7.5 Hz from MJD 51463 to 51473, i.e., within 10 days.  After that, QPOs did not reappear.

\subsection{XTE 1118+480}
We analyzed 2005 outburst of XTE 1118+480 and result is shown in Fig.~\ref{figxte}b. 
Data is from MJD 53383 (13/01/2005) to MJD 53396 (26/01/2005). During this outburst, 
CE was varying between $3$\% and $6$\%. In not too many days QPOs were observed during the outburst \citep{rem05}. 

\subsection{GRO J1655-40}
We analyzed 2005 outburst of GRO J1655-40 and result shown in Fig.~\ref{figxte}c. 
The compact object is analyzed from MJD 53435 (06/03/2005) to MJD 53632 (19/09/2005). 
During this outburst CE was varying between 2\% and 0.1\%. 
QPO variation is already reported in \citet{C08}.

\subsection{4U 1543-47}
We analyzed 2002 outburst of 4U 1543-47 and results are shown in 
Fig.~\ref{figxte}d. Compact object is analyzed from MJD 52442 (17/06/2002) to 
MJD 52480 (25/07/2002). Here CE varies between $0.3$\% to $0.01$\%.  
In case of this outburst, QPO is increased from $5$ to $8$ Hz from MJD 52442 to MJD 52460, 
then decreased to $1$ Hz from MJD 52460 to MJD 52480. 

\section{Summary of Results and a Comparison with GRS 1915+105}

In Table:~3, we summarize results of our analysis. Maximum and minimum values of 
CE, luminosity, QPO frequency, and mass of compact objects are put together. 
It is instructive to compare results with those obtained for a highly variable
black hole candidate GRS 1915+105 \citep{P11, P13} which is believed to be in a soft-intermediate 
state of some long duration outburst. Thus, Table:~3 also gives results 
of this source. Since CE is defined in a way to eliminate the effects of the 
mass, our comparison is meaningful, even when the mass varies by a factor of more than three.
In GRS 1915+105, CE varies strongly with a low value for generally softer classes, to a high value for 
generally harder classes. We clearly note that CE of GRS 1915+105 varies from $0.005$ to 
$0.8$, both ends being far from extreme values. These numbers indicate that if GRS 1915+105 
underwent an outburst long ago, it is still in a soft-intermediate state. The CE of GRS 1915+105 
was found to be very meaningful since occurrence of variability class transitions of this object 
follows the sequence of increasing or decreasing CE \citep{P13}.

\begin{table}
%\scriptsize{
\addtolength{\tabcolsep}{5.0pt}
 \label{symbols}
\begin{tabular}{@{}|c|c|c|c|c|c|c|c|c|}
%\multicolumn{12}{|c|}
%{Table 1: Data sets analyzed for all the compact objects.}\\
\hline 
Compact & Outburst & Mass & $CE_{min}$ & $CE_{max}$ & L$_{min}$ & L$_{max}$ & QPO$_{min}$ & QPO$_{max}$  \\ 
Object & (Year) & ($\approx M_\odot$) & ($\%$) & ($\%$) & (L$_{Edd}$) & (L$_{Edd}$) & (Hz) & (Hz) \\ 
\hline
XTE J1550-564 & 1998 & 9.6 & 0.0011&  2.78 & 0.26 & 1.8 & 0.3 & 12.7 \\
\hline
XTE J1550-564 & 2000 & 9.6 & 0.4 & 1.7 & 0.01 & 0.17 & 0.3 & 6.3 \\
\hline
GX 339-4 & 2007 & 7.5 & 0.02 & 0.4 & 0.006 & 0.26 & 0.6 & 10.0\\
\hline
GX 339-4 & 2010 & 7.5 & 0.03 & 0.375 & 0.007 & 0.16 & 0.3 & 6.0 \\   
\hline
H 1743-322 & 2008 & 10.0 & 0.1 & 0.3 & 0.01 & 0.04 & 1.0 & 5.0 \\
\hline
H 1743-322 & 2009 & 10.0 & 0.17 & 0.37 & 0.02  & 0.18 & 1.0 & 3.7 \\
\hline
XTE J1859+226 & 2000 & 4.5 & 0.004 & 0.05 & 0.01 & 0.20 &  3.0 & 7.6 \\
\hline
XTE J1118+480 & 2005 & 8.5 & 2.2 & 6.4 & 0.0003 & 0.001 & - & -\\
\hline
GRO J1655-40 & 2005 & 7.02 & 0.011 & 2.24 & 0.003 & 0.22 & 0.4 & 8.0 \\
\hline
4U 1543-47 & 2002 & 9.4 & 0.012 & 0.33 & 0.01 & 0.59 & 1.0 & 8.0 \\
\hline
GRS 1915+105 & - & 14 & 0.005 & 0.8 & 0.8 & 1.4 & 3.0 & 10.0 \\
\hline
%-& - &-  & --   \\ 
\end{tabular}
%}
\caption{Variation of CE, Luminosity and QPO for different compact objects during their outbursts. 
Data for GRS 1915+105 is presented for comparison.}
\end{table}

\section{Discussions \& Conclusions}

In this paper, we analyzed several black hole candidates which exhibit outbursts and 
shown the time dependence of Comptonizing efficiency, QPOs and the spectral index. 
We tried to understand the results using the TCAF model of \citep{C95} 
and its time varying form, namely, propagatory oscillating shock (POS) model of the 
outbursts \citep{C08, C09, D10, DD10}, though any model which relies on soft photons
and inverse Comptonization would be fine, since CE depends on the ratio of Comptonized photons and soft seed photons.
In TCAF model, a standard Keplerian disk is surrounded by a faster moving low-angular momentum flow 
(sub-Keplerian) which produces a centrifugal pressure supported shock where the flow is puffed up and 
produces the so-called Compton cloud to inverse Comptonize intercepted soft photons coming from the Keplerian disk. 
Oscillation of post-shock region or CENBOL causes low frequency QPOs in black hole candidates. 

In the backdrop of this model, our goal is to study how the optical depth of the CENBOL changes with time in 
a generic outburst source. If we had computed hardness ratio or HR, where the soft and hard photons 
are counted using the same fixed energy bins, a comparison of its behavior from one object to another would 
not be possible. This is because seed photons of one black hole could be Comptonized photon for another. On the 
other hand, CE as defined by us characterizes soft and hard photons objectively and as such does 
not depend on mass of the black hole or its accretion rate. Thus a comparison 
is possible. It is true that we restrict ourselves from 0.1 to 40 keV photons. 
However, for stellar mass black holes, these boundary values are far away from relevant energies in 
which respective photons are important.

We came to a conclusion that generally speaking, all outbursts start and end with a large 
Compton cloud size, though not necessarily with the highest optical depth. As the outburst progresses, 
CE becomes minimum at peak of an outburst i.e., size of CENBOL becomes very small. 
We find that CE in outburst sources may vary from almost $\sim 0.0$ to $\sim 3$\%. 
In contrast, variable source GRS 1915+105 
has CE between $0.005$ to $0.8$ and has relatively high luminosity (even after 
factoring out effects of mass of the black hole), suggesting that it is 
in a soft-intermediate state which usually appears after the peak of a possible outburst.
If so, in future, this source may slow down its activities when the viscosity in this system
is reduced as in any other outbursting source.

\begin{figure}
\begin{center}
\includegraphics[width=7.0truecm,angle=270]{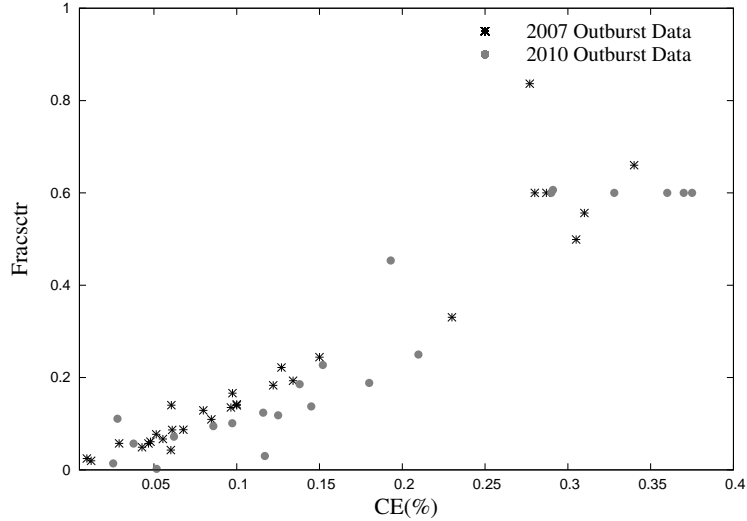}
\caption{Variation of fracsctr with CE(\%) for 2007 and 2010 outbursts of GX 339-4 data. 
Black stars and gray circle points respectively represent data of 2007 and 2010 outbursts. 
This plot shows variation of fracsctr with CE(\%) computed in 0.1-40 keV range.}
\label{figfrac}
\end{center}
\end{figure}

We would like to stress that CE computed by us counts soft photons from $0.1$ keV 
to an upper energy limit automatically detected by our fitting method. If we concentrated 
on photons in the observed $2-40$ keV range and counted the number of soft photons, 
we would have obtained a different, in fact, much higher CE. This is
because we would have grossly underestimated number of soft photons. 
In Fig.~\ref{figfrac}, we show a comparison of CE computed by our method with `fracsctr' 
obtained from {\it  simpl} model \citep{st11} for GX 339-4 data. 
{\it simpl} is an empirical model of Comptonization which computes fraction of  
photons from an input seed spectrum which are scattered into a power-law component. Fracsctr is 
fraction (in the scale up to 1.0) of input seed photons is Comptonized.
Black and gray points are for the 2007 and 2010 outburst data respectively. Black stars and gray circles
represent variations of fracsctr with CE when soft photons are taken from 0.1-40 keV CE(\%) range.
One generally sees that one increases with the other.
From Fig.~\ref{figfrac}, we can see that fracsctr value varies around 
0 to 0.8 this means 0-80 $\%$ photons are interacting with hot electron cloud. But if we 
compare with CE then only 0.005-0.35 $\%$ of photons are interacting with 
hot electron cloud which is more realistic. Computed CE values do support relative shape of 
CENBOL which is expected to intercept less than a few percent of soft photons from the Keplerian disk \citep{C95}.

\begin{figure}
\begin{center}
\includegraphics[width=7.0truecm,angle=0]{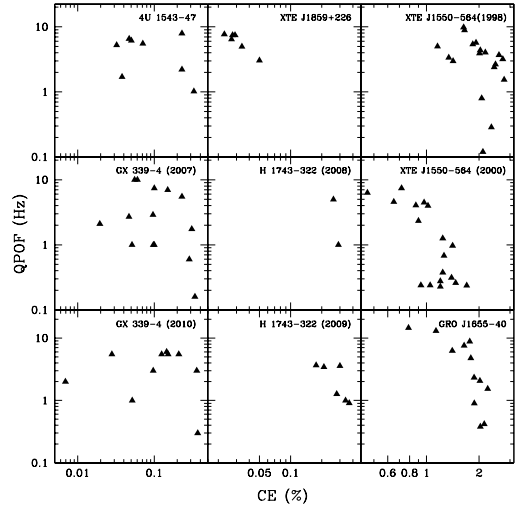}
\caption{Variation of QPO frequency with CE(\%) for all outbursts. The general trend is that 
the frequency increases as CE is decreased.} 
\label{qce}
\end{center}
\end{figure}

Another important issue is to correlate spectral and timing data. We mentioned that CE is higher for 
harder states. This is because location and height of shock wave are large. We have also 
mentioned that low frequency QPOs are due to oscillations of these shock waves and frequency is
inverse of the infall time scale from post-shock region \citep{CM00} till the inner sonic point. 
As such, larger shock radius means smaller QPO frequency, which is exactly what we observe! 
If we combine these results, then TCAF model would predict
that CE would be larger when the QPO frequency is smaller and vice versa. In Fig.~\ref{qce}, 
we show measured QPO frequencies as a function of our CE for all the outbursts discussed in this paper.
We find that CE is inversely correlated with QPO frequencies, although the relationship is not very tight
in some cases. This could be because of presence of radio jets observed in all these outburst sources 
\citep{BR02,Cas10,Cor10,K05,B04,Cor08, Mig07,H95,K03}. 
While CE is influenced by electron clouds at base of a jet (post-shock region in TCAF
solution), QPO frequency is not directly affected by the jet, but only by location of 
oscillating shock. This could be a primary reason
why some outbursters do not show a tight correlation. However, this aspect requires further investigation. 

\begin{figure}
\includegraphics[width=6.35truecm,angle=270]{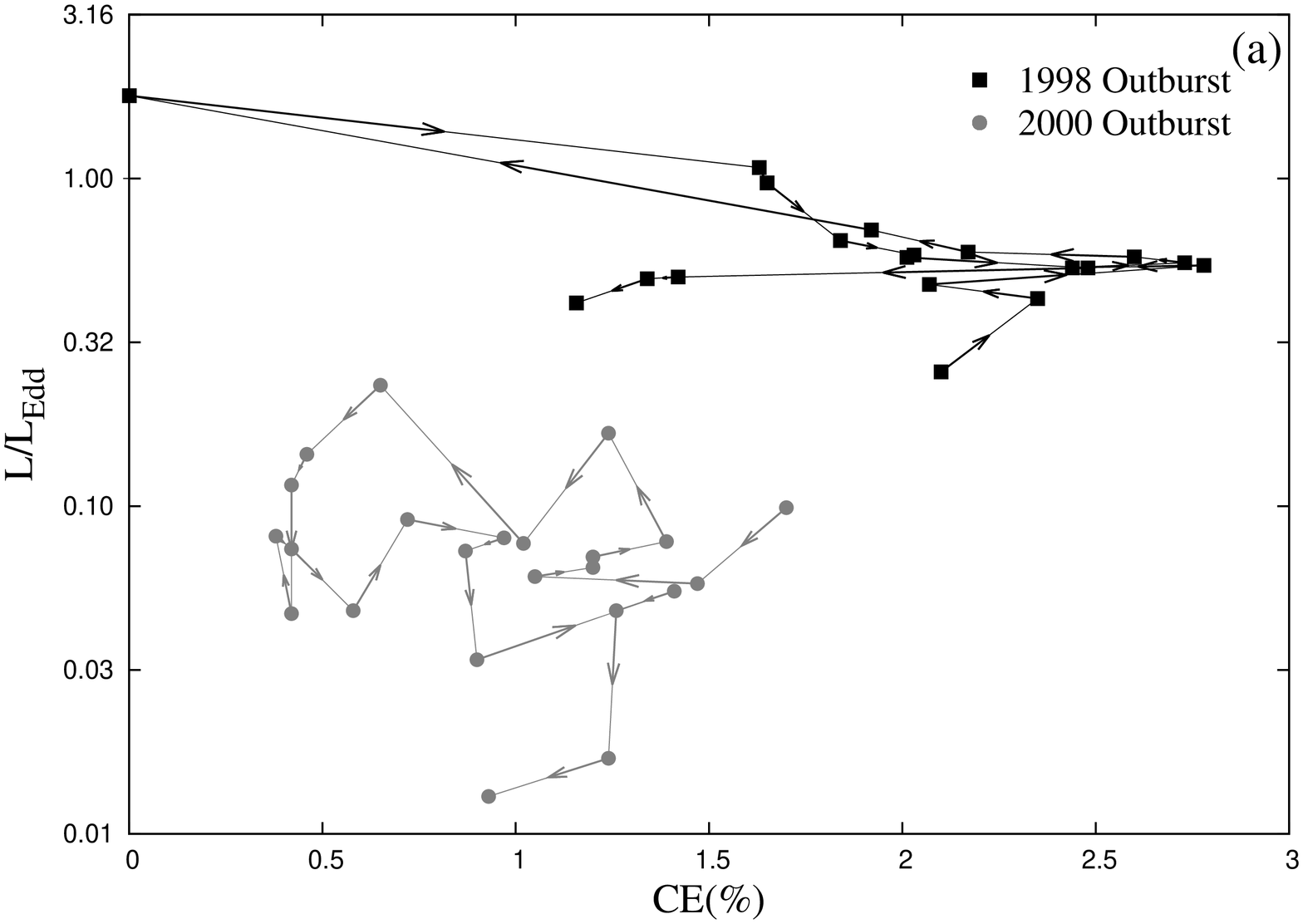}
\includegraphics[width=6.35truecm,angle=270]{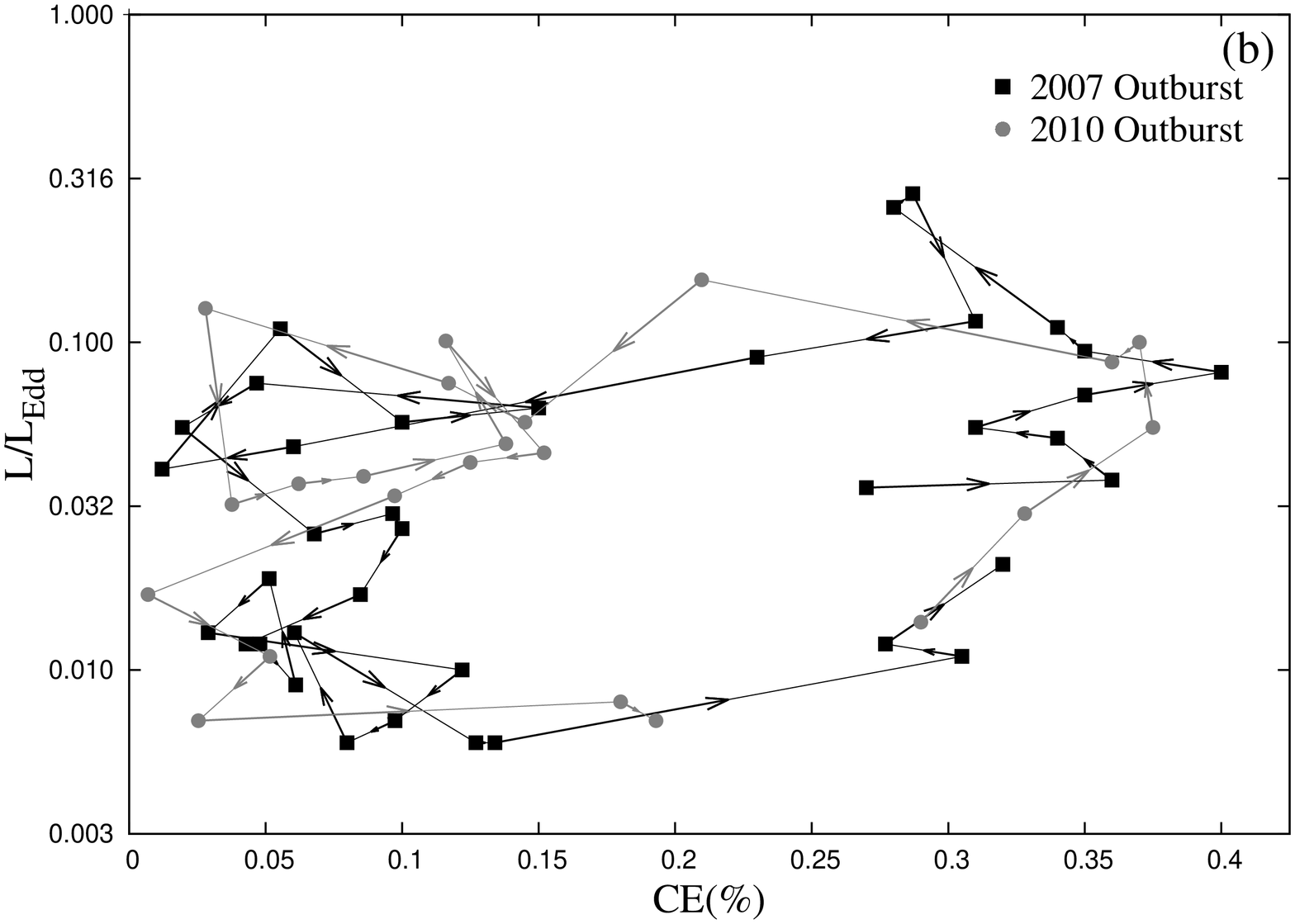}
\caption{Variation of CE(\%) with $(L/L_{Edd})$ for 
(a) 1998(black) and 2000(gray) outburst of XTE J1550-564,
(b) 2007(black) and 2010(gray) outburst of GX 339-4.}
\label{qdia}
\end{figure}

It is also interesting to study variation of CE with luminosity in analogy with hardness ratio-intensity 
plots in the literature \citep{f04}. Since hardness ratio is obtained from  
a fixed range of X-ray photons, independent of mass of black hole, 
it is difficult to interpret what a conventional colour-colour diagram really means. 
In Figs.~\ref{qdia}a, and \ref{qdia}b, we plot CE vs. luminosity for both outbursts of 
XTE 1550-564 and GX 339-4 respectively. 
In Fig.~\ref{qdia}(a), dark filled boxes represent variation of CE(\%) with natural
logarithm of luminosity in Eddington unit during 1998 outburst and gray filled circles represent
same for 2000 outburst. In Fig.~\ref{qdia}(b), black    
filled boxes represent 2007 outburst of GX 339-4, while gray filled circles represent the same 
for 2010 outburst of GX 339-4. For all plots, arrows of corresponding colour are provided to understand 
evolution of data during outbursts. Figures provide an idea about the
relation between geometric size of the electron cloud and total luminosity during 
outburst. In case of XTE 1550-564 (Fig.~\ref{qdia}a), we observe that both outbursts started and ended with high CEs, though
final value of CE is lower than its initial value. However, this could be due to 
incompleteness of observation during onset of the outburst. Outburst of 2000 is clearly 
weaker and short lived. 
In both the cases, there is a hysteresis effect in that paths in rising and declining phases
are not the same. This is expected, since formation time and disappearance time of a Keplerian 
disk by viscous effects are not identical \citep{G13}.  
In case of GX 339-4, both Figures are generally overlapping and thus both were of roughly equal strength.
We note that CE is not necessarily highest at beginning or end in all these cases, 
though it is close to highest values achieved during outburst. 
This may be because CE is sensitive to the optical depth and not just physical size of the CENBOL. This 
is also the reason why QPO frequency variation (sensitive to the size of CENBOL) is 
not tightly correlated with CE.
An important result is that CE of GX 339-4 is much lower as compared to that of XTE J1550-564. Since 
with spin the size of the CENBOL shrinks at least by a factor of two \citep{C96}, 
the degree of interception would also be reduced by the same factor. It is well known that spin of
XTE J1550-564 is moderate \citep{st11} while the spin of GX 339-4 is very high \citep{kol10}.
It is possible that we are seeing effects of large spin in GX 339-4 in our calculation also.

%WHAT IS THE POINT IS GIVING THIS PLOT? THEY LOOK IDENTICAL!
%In Fig.~\ref{q40} we have reproduced the Fig.~\ref{qdia}b. Here we have recalculated CE by integrating
%%soft photons from 0.1 to 40 keV. 
%
%\begin{figure}
%\includegraphics[width=5.00truecm,angle=270]{qdia_gx_40.ps}
%%\includegraphics[width=5.00truecm,angle=270]{qdia_gx_new.ps}
%\caption{Variation of CE(\%) with $(L/L_{Edd})$ for 2007(black) and 2010(gray) outburst 
%of GX 339-4. Here upto 40 keV soft photons are calculated.}
%\label{q40}
%\end{figure} 
%

In order to show that soft photons beyond $dbb_e$ really do not contribute significantly, we recomputed 
the above values with soft photons till $40$ keV. We do not find any significant difference in the result. 
So we believe that our results with dynamically obtained black body photon energy range yields 
sufficiently accurate results.

From the Table:~3, we see that range of CE is different for different objects. Duration of outbursts,
QPO frequency range etc. are also found to be different. There is also a considerable scatter
in CE which may be due to outflows from the CENBOL. All these require a more thorough analysis for 
a unifying understanding of the outbursts. This will be addressed in our future publications. 
Another way to improve the result is to use TCAF model itself to fit the spectra and obtain 
blackbody energy range more accurately. A third, and more serious point is to consider 
spin of black hole which will reduce CE by virtue of smaller CENBOL size for higher spins. 
These aspects will be looked into in future.

\section{Acknowledgment}
We acknowledge the referee for his useful advice to improve the paper.
PSP acknowledges SNBNCBS-PDRA Fellowship.

\end{document}